\documentclass[prd,twocolumn,superscriptaddress,nofootinbib,preprintnumbers]{revtex4}
\usepackage[latin1]{inputenc}
\usepackage{graphicx,color,amsmath,amsxtra}
\usepackage{epsf}
\usepackage{amssymb}
\usepackage{enumerate}
\usepackage{hhline}
\usepackage{array}
\usepackage{tabularx}

\newcommand{\bear}{\begin{array}}  \newcommand{\eear}{\end{array}}
\newcommand{\bea}{\begin{eqnarray}}  \newcommand{\eea}{\end{eqnarray}}
\newcommand{\beq}{\begin{equation}}  \newcommand{\eeq}{\end{equation}}
\newcommand{\bef}{\begin{figure}}  \newcommand{\eef}{\end{figure}}
\newcommand{\bec}{\begin{center}}  \newcommand{\eec}{\end{center}}
\newcommand{\rd}{{\rm d}}

\newcommand{\ti}{\tilde} 
\def\be{\begin{equation}}
\def\ee{\end{equation}}
\def\bea{\begin{eqnarray}}
\def\eea{\end{eqnarray}}
\def\beq{\begin{eqnarray}}
\def\eeq{\end{eqnarray}}

\def\be{\begin{equation}}
\def\ee{\end{equation}}
\def\bea{\begin{eqnarray}}
\def\eea{\end{eqnarray}}
\def\beq{\begin{eqnarray}}
\def\eeq{\end{eqnarray}}

\begin{document}

\title{Equilibrium thermodynamics in modified gravitational theories}

\author{Kazuharu Bamba}
\affiliation{Department of Physics, National Tsing Hua University, Hsinchu, 
Taiwan 300}
%\email{bamba@phys.nthu.edu.tw}

\author{Chao-Qiang Geng}
\affiliation{Department of Physics, National Tsing Hua University, Hsinchu, 
Taiwan 300}
%\email{geng@phys.nthu.edu.tw}

\author{Shinji Tsujikawa}
\affiliation{Department of Physics, Faculty of Science, Tokyo University of Science,
1-3, Kagurazaka, Shinjuku-ku, Tokyo 162-8601, Japan}
%\email{shinji@rs.kagu.tus.ac.jp}

%\date{\today}

%%%%%%%%%%%%%%%%%%%%%
%  Abstract
%%%%%%%%%%%%%%%%%%%%%
\begin{abstract}

We show that it is possible to obtain a picture of equilibrium thermodynamics 
on the apparent horizon in the expanding cosmological background
for a wide class of modified gravity theories with the Lagrangian 
density $f(R, \phi, X)$, where $R$ is the Ricci scalar and 
$X$ is the kinetic energy of a scalar field $\phi$.
This comes from a suitable definition of an energy momentum tensor
of the ``dark'' component that respects to a local energy conservation
in the Jordan frame.
In this framework the horizon entropy $S$ corresponding to  
equilibrium thermodynamics is equal to a quarter of the horizon area $A$
in units of gravitational constant $G$, as in Einstein gravity.
For a flat cosmological background with a decreasing Hubble parameter,
$S$ globally increases with time, as it happens for viable
$f(R)$ inflation and dark energy models. 
We also show that the equilibrium description in terms of 
the horizon entropy $S$ is convenient because it 
takes into account the contribution of both the horizon entropy 
$\hat{S}$ in non-equilibrium thermodynamics 
and an entropy production term.

\end{abstract}
%%%%%%%%%%%%%%%%%%%%%

%----------------------------
\pacs{
04.50.-h, 04.70.Dy, 95.36.+x, 98.80.-k}
%\pacs{
%Keywords:
%physics of the early universe
%}
%\preprint{}
%\hspace{13.0cm}
%----------------------------

\maketitle
%==============================================================================

%%%%%%%%%%%%%%%%%%%%%%%%%%%
%%%  (Sec. I)
%%%%%%%%%%%%%%%%%%%%%%%%%%%

\section{Introduction}

The discovery of black hole entropy by Bekenstein opened up 
a window for a profound physical connection between gravity 
and thermodynamics \cite{Beken}. 
The gravitational entropy $S$ in Einstein gravity is proportional 
to the horizon area $A$ of black holes, such that $S=A/(4G)$,
where $G$ is gravitational constant. 
A black hole with mass $M$ obeys the first law of
thermodynamics, $T \rd S=\rd M$ \cite{Bardeen}, 
where $T=|\kappa_s|/(2\pi)$ is a Hawking temperature 
determined by the surface
gravity $\kappa_s$ \cite{Hawking}. 

Since black hole solutions follow from Einstein field equations, 
the first law of black hole thermodynamics implies some 
connection between thermodynamics and Einstein equations.
In fact Jacobson \cite{Jacobson} showed that Einstein equations can 
be derived by using the Clausius relation $T\rd S=\rd Q$ on all local acceleration 
horizons in the Rindler space-time together with the relation $S \propto A$, 
where $\rd Q$ and $T$ are the energy flux across the horizon and
the Unruh temperature seen by an accelerating observer just inside 
the horizon, respectively.
This approach was applied to a number of cosmological 
settings such as quasi de Sitter inflationary universe \cite{Frolov,Danielsson}
and dark energy dominated universe \cite{Bousso}.

Unlike stationary black holes the expanding universe 
has a dynamically changing apparent horizon.
For dynamical black holes, Hayward \cite{Hayward} 
developed a framework to deal with 
the first law of thermodynamics on a trapping horizon 
in Einstein gravity (see Ref.~\cite{Paddy}
for related works). This was extended to the 
Friedmann-Lema\^{i}tre-Robertson-Walker (FLRW) 
space-time \cite{Akbar}, in which the Friedmann 
equation can be written in the form 
$T\rd S=-\rd E+W \rd V$, where 
$E$ is the intrinsic energy and $W$ is the work density 
present in the dynamical background.
For matter contents of the universe with energy density $\rho$
and pressure $P$, the work density is given by $W=(\rho-P)/2$.
Note that the energy flux $\rd Q$ in the Jacobson's formalism
is equivalent to $-\rd E+W \rd V$ in the FLRW background.

In the theories where the Lagrangian density $f$ is a non-linear
function in terms of the Ricci scalar $R$ 
(so called ``$f(R)$ gravity''), Eling {\it et al.} \cite{Eling} 
pointed out that a non-equilibrium treatment 
is required such that the Clausius relation is modified to 
$\rd \hat{S}=\rd Q/T+\rd_i \hat{S}$. Here the horizon entropy  is
defined by $\hat{S}=F(R)A/(4G)$ with $F(R)=\partial f/\partial R$
and $\rd_i \hat{S}$ describes a bulk viscosity entropy production term.
The variation of the quantity $F(R)$ gives rise to the non-equilibrium 
term $\rd_i \hat{S}$ that is absent in Einstein gravity, 
where a hat denotes the quantity in the non-equilibrium thermodynamics. 

The connections between thermodynamics and modified gravity 
have been extensively discussed, including the theories such as
$f(R)$ gravity \cite{Akbar2,Gong,Wu1,Wu2,Bamba09}, 
scalar-tensor theory \cite{Gong,Wu1,Wu2,Cai07}, 
Gauss-Bonnet and Lovelock gravity \cite{Cai07,Paddy2,CaiCao08}, 
and braneworld models \cite{braneth} 
(see also Ref.~\cite{Cai:2005ra}). 
In Gauss-Bonnet and Lovelock gravity, it is possible to 
rewrite Einstein equations in the form of equilibrium thermodynamics \cite{Cai07}.
On the other hand, in $f(R)$ gravity and scalar-tensor theories with 
the Lagrangian density $F(\phi)R$, 
it was shown in Refs.~\cite{Akbar2,Cai07,Wu1,Wu2} that, 
by employing the Hayward's dynamical framework, 
a non-equilibrium description of thermodynamics arises
for the Wald's horizon entropy defined 
by $\hat{S}=FA/(4G)$ \cite{Wald1,Wald2}.
This is consistent with the results obtained by Eling {\it et al.} \cite{Eling}.

The appearance of a non-equilibrium entropy production term 
$\rd_i \hat{S}$ is intimately related to the theories in which 
the derivative of the Lagrangian density $f$ with respect to 
$R$ is not constant.
It is of interest to see whether an equilibrium description of 
thermodynamics is possible in such modified gravity theories.
In this paper we will show that equilibrium thermodynamics exists 
for the general Lagrangian density $f(R, \phi, X)$, 
where $f$ is function of $R$, a scalar field $\phi$, 
and a field kinetic energy $X$.
This is possible by introducing the 
Bekenstein-Hawking entropy with a suitable redefinition of the
``dark component'' that respects a local energy conservation.
This approach is convenient because the horizon entropy $S$ defined 
in this framework involves the information of the horizon entropy $\hat{S}$ 
in non-equilibrium thermodynamics together with the 
entropy production term.

%%%%%%%%%%%%%%%%%%%%%%%%%%%%%%%%%%%%%%
\section{Thermodynamics in modified gravity--non-equilibrium picture}
\label{noneqsec}
%%%%%%%%%%%%%%%%%%%%%%%%%%%%%%%%%%%%%%

We start with the following action
\begin{equation}
I = \frac{1}{16\pi G} \int \rd^4 x \sqrt{-g} f(R,\phi,X)
-\int \rd^4 x\,{\mathcal{L}}_{M}
(g_{\mu \nu}, \Psi_M)\,,
\label{eq:2.1}
\end{equation}
where $g$ is the determinant of the metric tensor $g_{\mu\nu}$,
${\mathcal{L}}_{M}$ is the matter Lagrangian that depends on $g_{\mu \nu}$
and matter fields $\Psi_M$, and $X=-\left(1/2\right) g^{\mu\nu}
{\nabla}_{\mu}\phi {\nabla}_{\nu}\phi$ is the kinetic term 
of a scalar field $\phi$ (${\nabla}_{\mu}$ is the 
covariant derivative operator associated with $g_{\mu \nu}$).
The action~(\ref{eq:2.1}) can describe a wide variety of 
modified gravity theories, 
e.g., $f(R)$ gravity, Brans-Dicke theories, scalar-tensor theories,
and dilaton gravity, in addition to quintessence and k-essence.

{}From the action~(\ref{eq:2.1}), the gravitational field equation and 
the equation of motion for $\phi$ are derived as~\cite{Hwang}
\begin{eqnarray}
& &F G_{\mu\nu}=
8\pi G T_{\mu \nu}^{(M)}+\frac12 g_{\mu \nu}
(f-RF)+\nabla_{\mu}\nabla_{\nu}F \nonumber \\
& &\,\,\,\,\,\,\,\,\,\,\,\,\,\,\,\,\,\,\,\,\,\,\,
-g_{\mu \nu} \Box F+\frac12 f_{,X}
\nabla_{\mu} \phi \nabla_{\nu} \phi\,,
\label{eq:2.3d} \\
& & \frac{1}{\sqrt{-g}} \partial_{\mu} \left( f_{,X}
\sqrt{-g} g^{\mu \nu} \partial_{\nu} \phi \right)+f_{,\phi}=0\,,
\label{eq:2.3} 
\end{eqnarray}
where $F \equiv \partial f/\partial R$,
$f_{,X} \equiv \partial f/\partial X$,
$f_{,\phi} \equiv \partial f/\partial \phi$, 
$T_{\mu \nu}^{(M)}=(2/\sqrt{-g}) 
\delta {\cal L}_M/\delta g^{\mu \nu}$, and 
$G_{\mu\nu}=R_{\mu\nu}-\left(1/2\right)g_{\mu\nu}R$ 
is the Einstein tensor.
For the matter energy momentum tensor $T^{(M)}_{\mu \nu}$ 
we consider perfect fluids of ordinary matter (radiation and non-relativistic matter) 
with total energy density $\rho_f$ and pressure $P_f$.

We assume the 4-dimensional Friedmann-Lema\^{i}tre-Robertson-Walker (FLRW) 
space-time with the metric,
\begin{equation}
\rd s^2=h_{\alpha \beta} \rd x^{\alpha} \rd x^{\beta}
+\bar{r}^2 \rd \Omega^2\,,
\label{eq:2.5}
\end{equation}
where $\bar{r}=a(t)r$ and $x^0=t, x^1=r$ with the 
2-dimensional metric $h_{\alpha \beta}={\rm diag}(-1, a^2(t)/[1-Kr^2])$.
Here $a(t)$ is the scale factor, $K$ is the cosmic curvature, and 
$\rd \Omega^2$ is the metric of 2-dimensional sphere with unit radius. 
In the background (\ref{eq:2.5}) we obtain the following field 
equations from Eqs.~(\ref{eq:2.3d}) and (\ref{eq:2.3}):
\begin{eqnarray}
& &3F \left( H^2+K/a^2 \right) \nonumber \\
& &=f_{,X}X+
\frac12 (FR-f)-3H\dot{F}+8\pi G \rho_f\,,
\label{fri1} \\
& & -2F \left( \dot{H}-K/a^2 \right) \nonumber \\
& &=f_{,X}X+\ddot{F}-H\dot{F}+8\pi G (\rho_f+P_f)\,,
\label{fri2}\\
& & \frac{1}{a^3} \left( a^3 \dot{\phi} f_{,X} 
\right)^{\cdot}=f_{,\phi}\,,
\label{fri3}
\end{eqnarray} 
where a dot represents a derivative with respect to cosmic 
time $t$ and the scalar curvature is given by 
$R=6 (2H^2+\dot{H}+K/a^2)$.
The perfect fluid satisfies the continuity equation 
\begin{equation}
\dot{\rho}_f+3H(\rho_f+P_f)=0\,.
\label{rhof}
\end{equation}

Equations (\ref{fri1}) and (\ref{fri3}) can be written as
\begin{eqnarray}
& & H^2+\frac{K}{a^2}=\frac{8\pi G}{3F} 
\left( \hat{\rho}_d+\rho_f \right)\,,
\label{frire1} \\
& & \dot{H}-\frac{K}{a^2}=-\frac{4\pi G}{F}
\left( \hat{\rho}_d+\hat{P}_d+\rho_f+P_f \right)\,,
\label{frire2}
\end{eqnarray} 
where 
\begin{eqnarray}
& & \hat{\rho}_d \equiv \frac{1}{8\pi G} \left[ f_{,X}X
+\frac12 (FR-f)-3H \dot{F} \right]\,,
\label{rodef1} \\
& & \hat{P}_d \equiv \frac{1}{8\pi G}
\left[ \ddot{F}+2H \dot{F}-\frac12 (FR-f) \right]\,.
\label{pdef1}
\end{eqnarray} 
We use a hat to represent quantities in the non-equilibrium 
description of thermodynamics. 
Note that $\hat{\rho}_d$ and $\hat{P}_d$ originate from the 
energy-momentum tensor $\hat{T}_{\mu \nu}^{(d)}$ defined by 
\begin{eqnarray}
\hat{T}_{\mu \nu}^{(d)} &\equiv& \frac{1}{8\pi G}
\biggl[ \frac12 g_{\mu \nu}
(f-RF)+\nabla_{\mu}\nabla_{\nu}F
-g_{\mu \nu} \Box F \nonumber \\
& &\,\,\,\,\,\,\,\,\,\,\,\,\,\,\,\,+\frac12 f_{,X}
\nabla_{\mu} \phi \nabla_{\nu} \phi  \biggr]\,,
\label{Tmunud}
\end{eqnarray}
where the Einstein equation can be written as
\begin{equation}
G_{\mu \nu}=\frac{8\pi G}{F}
\left( \hat{T}_{\mu \nu}^{(d)}+
T_{\mu \nu}^{(M)} \right)\,.
\label{Ein1}
\end{equation}

If we define the density $\hat{\rho}_d$ and the pressure $\hat{P}_d$
of ``dark'' components in this way, we find that these obey the 
following equation 
\begin{equation}
\dot{\hat{\rho}}_d+3H(\hat{\rho}_d+\hat{P}_d)
=\frac{3}{8\pi G} (H^2+K/a^2) \dot{F}\,,
\label{rhocon1}
\end{equation}
where we have used Eq.~(\ref{fri3}).
For the theories with $\dot{F} \neq 0$ the r.h.s. of Eq.~(\ref{rhocon1})
does not vanish, so that the standard continuity equation does not hold.
This happens for $f(R)$ gravity and scalar-tensor theory.

Let us proceed to the thermodynamical property of the theories given above.
First of all the apparent horizon is determined by the condition 
$h^{\alpha \beta} \partial_{\alpha} \bar{r} \partial_{\beta} \bar{r}=0$, 
which means that the vector $\nabla \bar{r}$ is null on the surface
of the apparent horizon. For the FLRW space-time the radius of 
the apparent horizon is given by 
$\bar{r}_A=\left( H^2+K/a^2 \right)^{-1/2}$.
Taking the time derivative of this relation and 
using Eq.~(\ref{frire2}) it follows that 
\begin{equation}
\frac{F\rd \bar{r}_A}{4\pi G}=\bar{r}_A^3 H
\left( \hat{\rho}_d+\hat{P}_d+\rho_f+P_f \right)\rd t\,.
\label{Fdr}
\end{equation}

The Bekenstein-Hawking horizon entropy in the Einstein gravity 
is given by $S=A/(4G)$, where $A=4\pi \bar{r}_A^2$ is the
area of the apparent horizon \cite{Bardeen,Beken,Hawking}.
In the context of modified gravity theories, Wald introduced
a horizon entropy $\hat{S}$ associated with a Noether 
charge \cite{Wald1,Wald2}. 
The Wald entropy $\hat{S}$ is a local quantity defined 
in terms of quantities on the bifurcate Killing horizon. 
More specifically, it depends on the variation of the Lagrangian 
density of gravitational theories with respect to the Riemann tensor. 
This is equivalent to 
$\hat{S}=A/(4G_{\rm eff})$, where $G_{\rm eff}=G/F$
is the effective gravitational coupling \cite{Brustein}.
Using the Wald entropy 
\begin{equation}
\hat{S}=\frac{AF}{4G}\,,
\label{Sdef}
\end{equation}
together with Eq.~(\ref{Fdr}), we obtain
\begin{equation}
\frac{1}{2\pi \bar{r}_A} \rd \hat{S}=4\pi \bar{r}_A^3 H
\left( \hat{\rho}_d+\hat{P}_d+\rho_f+P_f \right)\rd t +
\frac{\bar{r}_A}{2G} \rd F\,.
\label{dSre}
\end{equation}
The apparent horizon has the following Hawking temperature 
$T=|\kappa_s|/(2\pi)$, where $\kappa_s$ is the surface gravity given by 
\begin{eqnarray}
\kappa_s &=& -\frac{1}{\bar{r}_A}
\left( 1-\frac{\dot{\bar{r}}_A}{2H\bar{r}_A} \right)
=-\frac{\bar{r}_A}{2} \left( \dot{H}+2H^2
+\frac{K}{a^2} \right) \nonumber \\
&=&
-\frac{2\pi G}{3F} \bar{r}_A 
\left( \hat{\rho}_T-3\hat{P}_T \right)\,,
\end{eqnarray}
with $\hat{\rho}_T=\hat{\rho}_d+\rho_f$ and $\hat{P}_T=\hat{P}_d+P_f$. 
As long as the total equation of state $w_T=\hat{P}_T/\hat{\rho}_T$
satisfies $w_T \le 1/3$ it follows that $\kappa_s \le 0$, which is the 
case for standard cosmology. Hence the horizon temperature is 
\begin{equation}
T=\frac{1}{2\pi \bar{r}_A}
\left( 1-\frac{\dot{\bar{r}}_A}{2H\bar{r}_A} \right)\,.
\label{tempe}
\end{equation}
Multiplying the term $1-\dot{\bar{r}}_A/(2H\bar{r}_A)$ for 
Eq.~(\ref{dSre}), we obtain 
\begin{eqnarray}
T \rd \hat{S}&=&4\pi \bar{r}_A^3 H (\hat{\rho}_d+
\hat{P}_d+\rho_f+P_f)\rd t
\nonumber \\
&&-2\pi  \bar{r}_A^2 (\hat{\rho}_d+\hat{P}_d+\rho_f+P_f)\rd 
\bar{r}_A+\frac{T}{G}\pi \bar{r}_A^2 \rd F. \nonumber \\
\label{TdS}
\end{eqnarray}

In Einstein gravity the Misner-Sharp energy \cite{Misner} is 
defined to be $E=\bar{r}_A/(2G)$.
In $f(R)$ gravity and scalar-tensor theory this was extended
to the form $\hat{E}=\bar{r}_AF/(2G)$ \cite{Gong}.
Using this latter expression for $f(R, \phi, X)$ theories,
it follows that 
\begin{equation}
\hat{E}=\frac{\bar{r}_A F}{2G}=
V \frac{3F (H^2+K/a^2)}{8\pi G}=V(\hat{\rho}_d+\rho_f)\,,
\label{Edef}
\end{equation}
where $V=4\pi \bar{r}_A^3/3$ is the volume inside
the apparent  horizon.
Using Eqs.~(\ref{rhof}) and (\ref{rhocon1}), we find 
the following relation 
\begin{eqnarray}
\rd \hat{E} &=& -4\pi \bar{r}_A^3 H (\hat{\rho}_d
+\hat{P}_d+\rho_f+P_f)\rd t
\nonumber \\
& &+4\pi \bar{r}_A^2 (\hat{\rho}_d+\rho_f) 
\rd \bar{r}_A+\frac{\bar{r}_A }{2G} \rd F\,.
\label{dE}
\end{eqnarray}

{}From Eqs.~(\ref{TdS}) and (\ref{dE}) we obtain
\begin{eqnarray}
T \rd \hat{S} &=& -\rd \hat{E}+2\pi \bar{r}_A^2 
(\hat{\rho}_d+\rho_f-\hat{P}_d-P_f) 
\rd \bar{r}_A \nonumber \\
& &+\frac{\bar{r}_A}{2G} 
\left( 1+2\pi \bar{r}_A T \right) \rd F\,.
\label{TS}
\end{eqnarray}
As in Refs.~\cite{Hayward,Cai07} we introduce the work density  
\begin{equation}
\hat{W}=(\hat{\rho}_d+\rho_f-\hat{P}_d-P_f)/2\,.
\label{work}
\end{equation}
Then Eq.~(\ref{TS}) reduces to
\begin{equation}
T \rd \hat{S}=-\rd \hat{E}+\hat{W} \rd V
+\frac{\bar{r}_A}{2G} \left( 1
+2\pi \bar{r}_A T \right) \rd F\,.
\end{equation}
This equation can be written in the form 
\begin{equation}
T \rd \hat{S}+T\rd_i \hat{S}=-\rd \hat{E}+\hat{W} \rd V\,,
\label{noneqfirst}
\end{equation}
where 
\begin{equation}
\rd_i \hat{S}=-\frac{1}{T} \frac{\bar{r}_A}{2G}
\left( 1+2\pi \bar{r}_A T \right) \rd F
=-\left( \frac{\hat{E}}{T}+\hat{S} \right) \frac{\rd F}{F}\,.
\label{diS}
\end{equation}

The new term $\rd_i \hat{S}$ may be interpreted as a term of 
entropy production in the non-equilibrium thermodynamics.
The theories with $F={\rm constant}$ lead to $\rd_i \hat{S}=0$, 
which means that the the first-law of the equilibrium 
thermodynamics holds. Meanwhile the theories with $\rd F \neq 0$, 
including $f(R)$ gravity and scalar-tensor theory, give rise to 
the additional term (\ref{diS}).

It is clear from Eq.~(\ref{rhocon1}) that the density $\hat{\rho}_d$ 
and the pressure $\hat{P}_d$ defined in Eqs.~(\ref{rodef1}) and 
(\ref{pdef1}) do not satisfy the 
standard continuity equation for $\dot{F} \neq 0$.
If it is possible to define $\hat{\rho}_d$ and $\hat{P}_d$ obeying the conserved 
equation, then we anticipate that the non-equilibrium description of 
thermodynamics may not be necessary.
In the next section we shall show that such a treatment is indeed possible.

%%%%%%%%%%%%%%%%%%%%%%%%%%%%%%%%%%%%%%%
\section{Equilibrium interpretation of thermodynamics in modified gravity}
\label{eqsec}
%%%%%%%%%%%%%%%%%%%%%%%%%%%%%%%%%%%%%%%

Let us consider equilibrium description of thermodynamics
for theories with the action (\ref{eq:2.1}).
Equations (\ref{fri1}) and (\ref{fri2}) can be written as 
\begin{eqnarray}
& & 3\left( H^2+\frac{K}{a^2} \right)=
8\pi G \left( \rho_d+\rho_f \right)\,,
\label{frire1d} \\
& & -2\left( \dot{H}-\frac{K}{a^2} \right)
=8\pi G \left( \rho_d+P_d+\rho_f+P_f \right),
\label{frire2d}
\end{eqnarray} 
where 
\begin{eqnarray}
\rho_d &\equiv& \frac{1}{8\pi G} \biggl[ f_{,X}X
+\frac12 (FR-f)-3H \dot{F}  \nonumber \\
& &~~~~~~~+3(1-F) (H^2+K/a^2) \biggr]\,,
\label{rodef1d}
\\
P_d &\equiv& \frac{1}{8\pi G}
\biggl[ \ddot{F}+2H \dot{F}-\frac12 (FR-f) \nonumber \\
& &-(1-F) (2\dot{H}+3H^2+K/a^2) \biggr]\,.
\label{pdef1d}
\end{eqnarray} 
If we define $\rho_d$ and $P_d$ in this way, they obey the 
following continuity equation
\begin{equation}
\dot{\rho}_d+3H (\rho_d+P_d)=0\,,
\label{conser}
\end{equation}
where we have used Eq.~(\ref{fri3}).
We then find that Eq.~(\ref{Fdr}) is replaced by 
\begin{equation}
\frac{\rd \bar{r}_A}{4\pi G}=\bar{r}_A^3 H
\left( \rho_d+P_d+\rho_f+P_f \right)\rd t\,.
\label{Fdr2}
\end{equation}

In the equilibrium description there is no need to define $G_{\rm eff}$ in the field equations 
in (\ref{frire1d}) and (\ref{frire2d}) in contrast to Eqs.~(\ref{frire1}) 
and (\ref{frire2}) in the non-equilibrium description. 
This originates from the redefinition of energy density $\rho_d$ and 
pressure $P_d$ of ``dark'' components defined in Eqs.~(\ref{rodef1d}) and 
(\ref{pdef1d}), respectively. This redefinition leads to the 
continuity equation (\ref{conser}). 
In other words, the energy-momentum 
conservation in terms of ``dark'' components is met. 
Since the perfect fluid of ordinary matter also satisfies the continuity 
equation (\ref{rhof}), the total energy density 
$\rho_T \equiv \rho_d+\rho_f$ 
and the total pressure $P_T \equiv P_d+P_f$ of the universe obey
the continuity equation 
\begin{equation}
\dot{\rho}_T + 3H (\rho_T+P_T)=0\,.
\label{C-E-T}
\end{equation}
Hence the equilibrium treatment of thermodynamics can 
be executed similarly to that in Einstein gravity. 
As a consequence, we introduce the Bekenstein-Hawking 
entropy \cite{Bardeen,Beken,Hawking}
\begin{equation}
S=\frac{A}{4G}=\frac{\pi}{G} 
\frac{1}{H^2+K/a^2} \,,
\label{Sdef2}
\end{equation}
unlike the Wald  entropy associated with $G_{\rm eff}$ in Eq.~(\ref{Sdef}) 
in the non-equilibrium thermodynamics. 
This allows us to obtain the equilibrium description of 
thermodynamics as that in Einstein gravity. 
Note that the Bekenstein-Hawking entropy (\ref{Sdef2}) is a global 
geometric quantity proportional to $A$, 
which is not directly affected by the difference of 
gravitational theories (i.e. by the difference of the quantity $F=\partial f/\partial R$).

{}From the definition (\ref{Sdef2}) it follows that 
\begin{equation}
\frac{1}{2\pi \bar{r}_A} \rd S=4\pi \bar{r}_A^3 H
\left( \rho_d+P_d+\rho_f+P_f \right)\rd t \,.
\label{dSre2}
\end{equation}
Using the horizon temperature given in Eq.~(\ref{tempe}), 
we obtain 
\begin{eqnarray}
T \rd S &=& 4\pi \bar{r}_A^3 H (\rho_d+P_d+\rho_f+P_f)\rd t
\nonumber \\
& &-2\pi  \bar{r}_A^2 (\rho_d+P_d+\rho_f+P_f)
\rd \bar{r}_A\,.
\label{TdS2}
\end{eqnarray}
Defining the Misner-Sharp energy to be 
\begin{equation}
E=\frac{\bar{r}_A}{2G}=
V(\rho_d+\rho_f)\,,
\label{Misner2}
\end{equation}
we find
\begin{equation}
\rd E=-4\pi \bar{r}_A^3 H (\rho_d+P_d+\rho_f+P_f)\rd t
+4\pi \bar{r}_A^2 (\rho_d+\rho_f) 
\rd \bar{r}_A\,.
\label{dE2}
\end{equation}
Due to the conservation equation (\ref{conser}), 
the r.h.s. of Eq.~(\ref{dE2}) does not include an additional term 
proportional to $\rd F$.
Combing Eqs.~(\ref{TdS2}) and (\ref{dE2}) gives
\begin{equation}
T \rd S=-\rd E+W \rd V\,,
\label{first}
\end{equation}
where the work density $W$ is defined by 
\begin{equation}
W=\left( \rho_d+\rho_f-P_d-P_f \right)/2\,.
\label{work2}
\end{equation}
Equation (\ref{first}) corresponds to the first law of
equilibrium thermodynamics. This shows that 
the equilibrium form of thermodynamics can be derived
by introducing the density $\rho_d$ and the pressure 
$P_d$ in a suitable way.

Plugging Eqs.~(\ref{Misner2}) and (\ref{work2}) into 
Eq.~(\ref{first}), we find 
\begin{equation}
T \dot{S}=V \left( 3H-\frac{\dot{V}}{2V} \right)
(\rho_d+\rho_f+P_d+P_f)\,.
\end{equation}
Using $V=4\pi \bar{r}_A^3/3$ and Eq.~(\ref{tempe}), 
it follows that 
\begin{eqnarray}
\dot{S} &=& 
6\pi H V \bar{r}_A(\rho_d+\rho_f+P_d+P_f) \nonumber \\
&=& -\frac{2\pi}{G} \frac{H (\dot{H}-K/a^2)}
{(H^2+K/a^2)^2}\,.
\label{dSre3}
\end{eqnarray}
The horizon entropy increases as long as the null energy condition 
$\rho_T+P_T = \rho_d+\rho_f+P_d+P_f \ge 0$ is satisfied. 
$S$ decreases for the total equation of state 
$w_T \equiv P_T/\rho_T<-1$, as it happens in General Relativity.
Meanwhile the Wald entropy (\ref{Sdef}) does not in general 
possess this property, as we will see for $f(R)$ inflation models 
in Sec.~\ref{infsec}.

The above equilibrium picture of thermodynamics is intimately 
related with the fact that there is an energy momentum tensor 
$T_{\mu \nu}^{(d)}$ satisfying the local conservation law
$\nabla^{\mu} T_{\mu \nu}^{(d)}=0$.
This corresponds to writing the Einstein equation in the form 
\begin{equation}
G_{\mu \nu}=8\pi G 
\left( T_{\mu \nu}^{(d)}+
T_{\mu \nu}^{(M)} \right)\,,
\label{redeein}
\end{equation}
where 
\begin{eqnarray}
T_{\mu \nu}^{(d)} &\equiv& \frac{1}{8\pi G}
\biggl[ \frac12 g_{\mu \nu}
(f-R)+\nabla_{\mu}\nabla_{\nu}F
-g_{\mu \nu} \Box F \nonumber \\
& &~~~~~~~~~~~~~~~~~~~~~~+\frac12 f_{,X}
\nabla_{\mu} \phi \nabla_{\nu} \phi+
(1-F)R_{\mu \nu} \biggr]\,.
\label{eneequ}
\end{eqnarray}
Defining $T_{\mu \nu}^{(d)}$ in this way, 
the local conservation of $T_{\mu \nu}^{(d)}$ follows from 
Eq.~(\ref{redeein}) because of the relations 
$\nabla^{\mu}G_{\mu \nu}=0$ and
$\nabla^{\mu}T_{\mu \nu}^{(M)}=0$.

One can show that the horizon entropy $S$ in the equilibrium 
picture has the following relation with $\hat{S}$ in the 
non-equilibrium picture:
\begin{eqnarray}
\rd S &=& \rd \hat{S} + \rd_i \hat{S}
+\frac{\bar{r}_A}{2GT}\rd F \nonumber \\
& &-\frac{2\pi (1-F)}{G}
\frac{H (\dot{H}-K/a^2)}{(H^2+K/a^2)^2}\,\rd t.
\label{Tre}
\end{eqnarray}
Using the relations (\ref{diS}) and (\ref{dSre2}),
Eq.~(\ref{Tre}) reduces to the following form 
\begin{equation}
\rd S=\frac{1}{F} \rd \hat{S}+\frac{1}{F}
\frac{2H^2+\dot{H}+K/a^2}
{4H^2+\dot{H}+3K/a^2}\,\rd_i \hat{S}\,,
\label{usere}
\end{equation}
where 
\begin{equation}
\rd_i \hat{S}=-\frac{6\pi}{G} \frac{4H^2+\dot{H}+3K/a^2}
{H^2+K/a^2} \frac{\rd F}{R}\,.
\end{equation}
While $S$ is identical to $\hat{S}$ in Einstein gravity ($F=1$),
the difference appears in modified gravity theories with $\rd F \neq 0$.
Equation (\ref{usere}) shows that the change of the horizon entropy $S$ 
in the equilibrium framework involves the information of 
both $\rd \hat{S}$ and $\rd_i \hat{S}$ in the non-equilibrium framework.

%%%%%%%%%%%%%%%%%%%%%%%%%%%
\section{Einstein frame in scalar-tensor theories}
\label{einsteinsec}
%%%%%%%%%%%%%%%%%%%%%%%%%%%

For some specific theories, the action (\ref{eq:2.1})
can be transformed to the so-called Einstein frame action
via the conformal transformation.
For example let us consider the following 
scalar-tensor theories with the action 
\begin{eqnarray}
I  &=& \int \rd^4 x \sqrt{-g} 
\left[ \frac{F(\phi)}{2\kappa^2}R+\omega(\phi)X-V(\phi)
\right] \nonumber \\
& &-\int \rd^4 x\,{\mathcal{L}}_{M}
(g_{\mu \nu}, \Psi_M)\,,
\label{stensor}
\end{eqnarray}
where $\kappa^2 \equiv 8\pi G$, and $F(\phi)$, $\omega(\phi)$, 
$V(\phi)$ are functions of $\phi$.
Under the conformal transformation, $\tilde{g}_{\mu \nu}=F g_{\mu \nu}$, 
we obtain the following action in the Einstein frame \cite{Maeda,Tsuji08}:
\begin{eqnarray}
I_E  &=& \int \rd^4 x \sqrt{-\ti{g}} 
\left[ \frac{1}{2\kappa^2}\ti{R}-\frac12 (\ti{\nabla}\varphi)^2
-U(\varphi) \right] \nonumber \\
& &-\int \rd^4 x\,{\mathcal{L}}_{M}
(F(\phi)^{-1}\ti{g}_{\mu \nu}, \Psi_M)\,,
\label{stensore}
\end{eqnarray}
where a tilde represents quantities in the Einstein frame, and 
\begin{equation}
\varphi \equiv \int \rd \phi \sqrt{\frac32 \left( \frac{F_{,\phi}}
{\kappa F} \right)^2+\frac{\omega}{F}}\,,\qquad
U=\frac{V}{F^2}\,.
\label{varphidef}
\end{equation}
Varying the action (\ref{stensore}) with respect to $\varphi$, we obtain 
the field equation 
\begin{equation}
\square \varphi-U_{,\varphi}-\frac{1}{\sqrt{-\ti{g}}}
\frac{\partial {\cal L}_M}{\partial \varphi}=0\,.
\label{fieldeq}
\end{equation}

In the FLRW background the following relations hold
\begin{equation}
\rd \ti{t}=\sqrt{F}\,\rd t\,,\qquad
\ti{a}=\sqrt{F}a\,.
\label{tare}
\end{equation}
Using the relation $-\partial {\cal L}_M/\partial \varphi=\sqrt{-\ti{g}}
\kappa Q (\varphi) \ti{T}_M$, where $Q(\varphi) \equiv -F_{,\varphi}/(2\kappa F)$
and $\ti{T}_M \equiv -\ti{\rho}_f+3\ti{P}_f$, the field 
equation (\ref{fieldeq}) reduces to 
\begin{equation}
\ddot{\varphi}+3 \ti{H} \dot{\varphi}+U_{,\varphi}
=-Q(\varphi) ( \ti{\rho}_f-3\ti{P}_f )\,.
\label{fieldeqva}
\end{equation}
In this section a dot represents a derivative with respect to $\ti{t}$.
Introducing $\ti{\rho}_{\varphi} \equiv \dot{\varphi}^2/2+U(\varphi)$ and 
$\ti{P}_{\varphi} \equiv \dot{\varphi}^2/2-U(\varphi)$, Eq.~(\ref{fieldeqva})
can be written as 
\begin{equation}
\dot{\ti{\rho}}_{\varphi}+
3\ti{H} (\ti{\rho}_{\varphi}+\ti{P}_{\varphi})=
-Q(\varphi) ( \ti{\rho}_f-3\ti{P}_f )
\dot{\varphi}\,.
\label{fieldeq2}
\end{equation}
Note that $\ti{\rho}_f$ and $\ti{P}_f$ in the Einstein frame 
are related with $\rho_f$ and $P_f$ in the Jordan frame via 
$\rho_f=F^2 \ti{\rho}_f$ and $P_f=F^2 \ti{P}_f$.
Using Eqs.~(\ref{rhof}) and (\ref{tare}), it then follows that 
\begin{equation}
\dot{\ti{\rho}}_f+
3\ti{H} (\ti{\rho}_f+\ti{P}_f)=
+Q(\varphi) ( \ti{\rho}_f-3\ti{P}_f )
\dot{\varphi}\,.
\label{rhoeq}
\end{equation}

Equations (\ref{fieldeq2}) and (\ref{rhoeq}) show that the field $\varphi$
is coupled to matter with the coupling $Q(\varphi)$ except 
for radiation \cite{Amencoupled}.
In Brans-Dicke theory with $F(\phi)=\kappa\phi$ and 
$\omega(\phi)=\omega_{\rm BD}/(\kappa\phi)$ ($\omega_{\rm BD}$ is 
a constant parameter), the coupling 
$Q$ is a constant \cite{Tsuji08}.
The $f(R)$ gravity in the metric formalism corresponds to 
$\omega_{\rm BD}=0$ and $V=(FR-f)/(2\kappa^2)$, in which case
$\kappa \varphi=\sqrt{3/2}\,\ln \,F$ from Eq.~(\ref{varphidef}).
Hence the coupling $Q(\varphi)=-F_{,\varphi}/(2\kappa F)$ 
is also a constant ($Q=-1/\sqrt{6}$) in metric $f(R)$ 
gravity \cite{APT}. 

{}From Eqs.~(\ref{fieldeq2}) and (\ref{rhoeq}) the total energy 
density $\ti{\rho}_T=\ti{\rho}_{\varphi}+\ti{\rho}_f$ and the 
total pressure $\ti{P}_T=\ti{P}_{\varphi}+\ti{P}_f$
satisfy the continuity equation, 
$\rd \ti{\rho}_T/\rd \ti{t}+3\ti{H} (\ti{\rho}_T+\ti{P}_T)=0$.
The following equations also hold in the Einstein frame
\begin{eqnarray}
& & 3(\ti{H}^2+K/\ti{a}^2)=\kappa^2 
\left( \ti{\rho}_{\varphi}+\ti{\rho}_f \right)\,,
\label{frie1} \\
& & 2(\dot{\ti{H}}-K/\ti{a}^2)=-\kappa^2
( \ti{\rho}_{\varphi}+\ti{P}_{\varphi}
+\ti{\rho}_f+\ti{P}_f )\,.
\label{frie2}
\end{eqnarray} 
We define several thermodynamical quantities
\begin{eqnarray}
 & &\ti{T}=\frac{|\ti{\kappa}_s|}{2\pi},\quad \ti{S}=\frac{\ti{A}}{4G},\quad
\ti{E}=\frac{\ti{\bar{r}}_A}{2G},\quad \ti{V}=4\pi \ti{\bar{r}}_A^3/3\,,\nonumber \\
& & \ti{W}=(\ti{\rho}_{\varphi}+\ti{\rho}_f-\ti{P}_{\varphi}-\ti{P}_f)/2\,,
\end{eqnarray}
where $\ti{\bar{r}}_A=(\ti{H}^2+K/\ti{a}^2)^{-1/2}$, $\ti{A}=4 \pi \ti{\bar{r}}_A^2$, 
and $\ti{\kappa}_s=-[1-(\rd \ti{\bar{r}}_A/\rd \ti{t})/(2\ti{H}\ti{\bar{r}}_A)]/\ti{\bar{r}}_A$.
Following the similar procedure as in Sec.~\ref{eqsec}, 
we arrive at the first law of thermodynamics, 
\begin{equation}
\ti{T}\rd \ti{S}=-\rd \ti{E}+\ti{W} \rd \ti{V}\,. 
\end{equation}
Hence the equilibrium description of thermodynamics
holds in the Einstein frame.

In the following we shall find the relation of the equilibrium description in the 
Jordan frame for the flat universe ($K=0$).
The horizon entropy and the Misner-Sharp energy in the Einstein frame 
are given, respectively, by 
\begin{equation}
\ti{S}=\frac{\pi}{G}\frac{1}{\ti{H}^2}\,,\qquad
\ti{E}=\frac{1}{2G}\frac{1}{\ti{H}}\,.
\end{equation}
These are different from those in the equilibrium picture in the Jordan 
frame ($S=\pi/(GH^2)$ and $E=1/(2GH)$) because of the difference 
of the Hubble parameter:
\begin{equation}
H=\sqrt{F} \left( \ti{H}-\frac{1}{2F} \frac{\rd F}{\rd \ti{t}} \right)
=\sqrt{F} \left( \ti{H}+\kappa Q \frac{\rd \varphi}{\rd \ti{t}}
\right)\,.
\label{Hrera}
\end{equation}
In the Jordan frame the following relations hold
\begin{eqnarray}
& &\rd E=-\frac{\rd H}{2GH^2},\quad
W \rd V=\left(3+\frac{1}{H^2}\frac{\rd H}{\rd t} \right)\rd E,\nonumber \\
& & T\rd S=\left( 2+\frac{1}{H^2}\frac{\rd H}{\rd t}  \right)\rd E.
\end{eqnarray}
Note that similar relations hold in the Einstein frame by adding 
the tilde for corresponding quantities.
Using Eq.~(\ref{Hrera}) we obtain 
\begin{equation}
\rd E=(\mu/\sqrt{F}) \rd \ti{E}\,,
\end{equation}
where 
\begin{equation}
\mu \equiv \frac{1+\kappa Q[ \ddot{\varphi}+(Q_{,\varphi}/Q-\kappa Q)
\dot{\varphi}^2-\ti{H}\dot{\varphi}]/\dot{\ti{H}}}
{(1+\kappa Q \dot{\varphi}/\ti{H})^2}\,, 
\label{mueq}
\end{equation}
and 
\begin{eqnarray}
& & W \rd V=\frac{\mu (3+\mu \dot{\ti{H}}/\ti{H}^2)}
{\sqrt{F}(3+\dot{\ti{H}}/\ti{H}^2)} \ti{W} \rd \ti{V}\,,\\
& & T\rd S=\frac{\mu (2+\mu \dot{\ti{H}}/\ti{H}^2)}
{\sqrt{F}(2+\dot{\ti{H}}/\ti{H}^2)} \ti{T} \rd \ti{S}\,.
\end{eqnarray}
Recall that the field $\varphi$ in Eq.~(\ref{mueq}) satisfies 
Eq.~(\ref{fieldeqva}).

Since $F=1$ and $\mu=1$ in Einstein gravity, one has 
$\rd E=\rd \ti{E}$, $W \rd V=\ti{W} \rd \ti{V}$, and 
$T \rd S=\ti{T} \rd \ti{S}$.
In scalar-tensor theories in which $F$ and $\mu$ dynamically 
change in time, the equilibrium description of thermodynamics 
in the Jordan frame is not identical to that in the Einstein frame.
We note that general modified gravity theories with the action 
(\ref{eq:2.1}) do not necessarily have the action in the Einstein 
frame. Even for such general theories we have shown 
in Sec.~\ref{eqsec} that the equilibrium picture of 
thermodynamics is present without any reference to 
the Einstein frame.

Moreover, we regard that the frame in which the baryons obey 
the standard continuity equation $\rho_m \propto a^{-3}$, 
i.e. the Jordan frame, is the ``physical'' frame where
physical quantities are compared with observations 
and experiments. The direct construction of the equilibrium 
thermodynamics in the Jordan frame is not only versatile but 
is physically well motivated.
Only for the theories in which the action in the Einstein 
frame exists we can find the relation between the 
thermodynamical quantities in the two frames, 
as we have done above.

%%%%%%%%%%%%%%%%%%%%%%
\section{Application to $f(R)$ theories}
\label{applysec}
%%%%%%%%%%%%%%%%%%%%%%

In this section we apply the formulas of the horizon entropies
in the Jordan frame to inflation and dark energy in $f(R)$ theories.
In particular the evolution of $S$ and 
$\hat{S}$ will be discussed during inflation (and reheating)
in $f(R)$ theories. 
We also study how the horizon entropies evolve
during an epoch of the late-time cosmic acceleration 
in $f(R)$ dark energy models.
In the following we assume the flat FLRW space-time ($K=0$). 

\subsection{Inflation}
\label{infsec}

It is known that cosmological inflation can be realized by the 
Lagrangian density of the form $f(R)=R+\alpha R^n$ ($\alpha, n>0$).
The first inflation model proposed by Starobinsky corresponds 
to $n=2$ \cite{Star80}. Let us consider the model $f(R)=R+\alpha R^n$
in the region $F=\rd f/\rd R=1+n \alpha R^{n-1} \gg 1$.

During inflation one can use the approximations $|\dot{H}/H^2| \ll 1$ 
and $|\ddot{H}/(H\dot{H})| \ll 1$. 
With these approximations, 
in the absence of matter fluids Eq.~(\ref{frire1d}) reduces to 
\begin{equation}
\frac{\dot{H}}{H^2}=-\beta\,,\qquad
\beta \equiv \frac{2-n}{(n-1)(2n-1)}\,.
\end{equation}
This gives the power-law evolution of the scale factor 
($a \propto t^{1/\beta}$), which means that inflation occurs
for $\beta<1$, i.e. $n>(1+\sqrt{3})/2$.
When $n=2$ one has $\beta=0$, so that $H$ is constant 
in the regime $F \gg 1$. The models with $n>2$ lead to 
the super-inflation characterized by $\dot{H}>0$ and 
$a \propto |t_0-t|^{-1/|\beta|}$ ($t_0$ is a constant).

The standard inflation with decreasing $H$ occurs for $0<\beta<1$, 
i.e. $(1+\sqrt{3})/2<n<2$. In this case the horizon entropy 
(\ref{Sdef2}) in the equilibrium framework
grows as $S \propto H^{-2} \propto t^2$ during inflation.
Meanwhile the horizon entropy $\hat{S}=F(R)A/(4G)$
in the non-equilibrium framework has a dependence 
$\hat{S} \propto R^{n-1}/H^2 \propto H^{2(n-2)} \propto
t^{2(2-n)}$ in the regime $F \gg 1$.
Hence $\hat{S}$ grows more slowly relative to $S$. 
This property can be understood from Eq.~(\ref{usere}), i.e.
\begin{equation}
\frac{\rd S}{\rd t}=\frac{1}{F} \frac{\rd \hat{S}}{\rd t}+
\frac{1}{F} \frac{2-\beta}{4-\beta} 
\frac{\rd_i \hat{S}}{\rd t}\,,
\label{dSrein}
\end{equation}
where 
\begin{equation}
\frac{\rd_i \hat{S}}{\rd t}=
\frac{12\pi \beta (4-\beta)}{G} HF_{,R}\,.
\label{hatSd}
\end{equation}
Here $F_{,R} \equiv \rd F/\rd R$. 
For the above model the term $F=n \alpha R^{n-1}$ 
evolves as $F \propto t^{2(1-n)}$. 
This means that $(1/F) \rd \hat{S}/\rd t \propto t$ in Eq.~(\ref{dSrein}), 
which has the same dependence as the time-derivative of $S$, i.e.
$\rd S/\rd t \propto t$.
The r.h.s. of Eq.~(\ref{hatSd}) is positive because $F_{,R}>0$ and $\beta>0$, 
so that $\rd_i \hat{S}/\rd t>0$.
We have $\rd_i \hat{S}/\rd t \propto t^{3-2n}$ and hence 
the last term on the r.h.s. of Eq.~(\ref{dSrein}) also grows 
in proportion to $t$.
Thus $S$ evolves differently from $\hat{S}$
because of the presence of the term $1/F$.

More precisely each term in Eq.~(\ref{dSrein}) is given by 
\begin{eqnarray}
\frac{\rd S}{\rd t} &=&\frac{2\pi \beta}{G}\frac{1}{H}\,,
\label{ent1}
\\
\frac{1}{F}\frac{\rd \hat{S}}{\rd t} &=&
\frac{2\pi \beta (2-n)}{G}\frac{1}{H}\,,
\label{ent2}
\\
\frac{1}{F} \frac{2-\beta}{4-\beta} 
\frac{\rd_i \hat{S}}{\rd t} &=& \frac{2\pi \beta (n-1)}
{G}\frac{1}{H}\,,
\label{ent3}
\label{each}
\end{eqnarray}
where we have used $F \gg 1$. 
When $(1+\sqrt{3})/2<n<2$, i.e. $0<\beta<1$, it follows 
that $\rd S>0, \rd \hat{S}>0, \rd_i \hat{S}>0$ for $\rd t>0$.
In the limit that $n \to 2$ the ratio $r=(2-n)/(n-1)$ 
of the r.h.s. of Eq.~(\ref{ent2}) to the r.h.s. of Eq.~(\ref{ent3})
approaches 0, so that the entropy production term $\rd_i \hat{S}$
gives a dominant contribution to $\rd S$.
When $n>2$, i.e. $\beta<0$, we have 
$\rd S<0, \rd \hat{S}>0, \rd_i \hat{S}<0$ for $\rd t>0$.
Hence the decrease of $S$ comes from the negative entropy 
production term $\rd_i \hat{S}$.
The entropy $S$ in the equilibrium framework
decreases for the theories with $\dot{H}>0$, whereas
$\hat{S}$ in the non-equilibrium framework can grow 
even in such cases unless the entropy production term
is taken into account.
The equilibrium description of thermodynamics allows us
to introduce the Bekenstein-Hawking entropy that mimics
the property in General Relativity.

%%%%%%%%%%%%%%
\begin{figure}
\includegraphics[height=3.4in,width=3.4in]{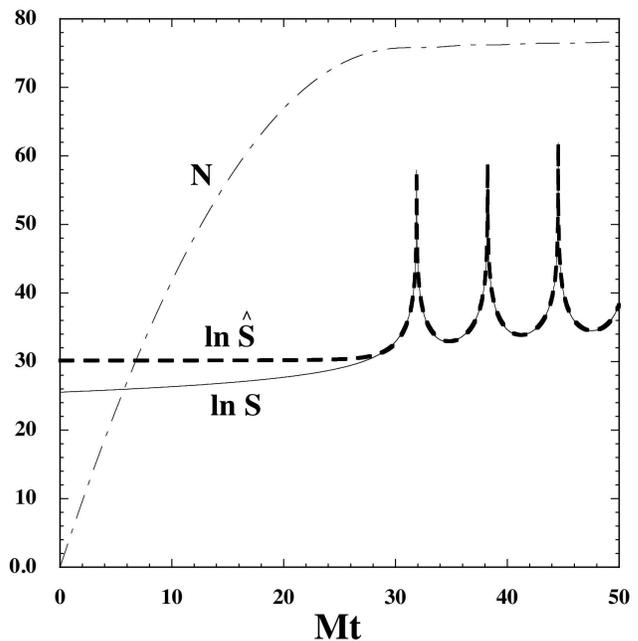}
\caption{\label{inflationfig} 
The variation of the horizon entropy $S=A/(4G)$ in the equilibrium picture 
as well as $\hat{S}=FA/(4G)$ in the non-equilibrium picture
for the inflation model $f(R)=R+R^2/(6M^2)$. We also show the evolution of the 
number of e-foldings $N=\ln (a/a_i)$ from the beginning of inflation. 
Initial conditions are chosen to be $R/M^2=305$ and $\dot{R}=0$.
The entropy $S$ evolves faster than $\hat{S}$ during inflation ($0<Mt<30$), 
but their evolution is similar after inflation in which $F$ is close to 1.}
\end{figure}
%%%%%%%%%%%%%%

In the Starobinsky's model $f(R)=R+R^2/(6M^2)$,  
the presence of the linear term in $R$ 
eventually causes inflation to end.
Without neglecting this linear term, we obtain the 
following equations: 
\begin{eqnarray}
& & \ddot{H}-\frac{\dot{H}^2}{2H}+3H\dot{H}
+\frac12 M^2 H=0\,,
\label{steq1}
\\
& & \ddot{R}+3H\dot{R}+M^2 R=0\,.
\label{steq2}
\end{eqnarray}
During inflation the first two terms in Eq.~(\ref{steq1})
can be neglected relative to others, which gives
$\dot{H} \simeq -M^2/6$.
We then obtain the solution $H \simeq H_i-(M^2/6)(t-t_i)$
with the Ricci scalar $R \simeq 12H^2-M^2$, 
where $H_i$ is the Hubble parameter at the onset of 
inflation (at $t=t_i)$. The accelerated expansion ends when 
the slow-roll parameter $\epsilon \equiv -\dot{H}/H^2 \simeq 
M^2/(6H^2)$ grows to the order of unity, i.e. $H\simeq M/\sqrt{6}$.
The horizon entropy $S$ grows as $S \propto \left[ H_i-(M^2/6)(t-t_i) \right]^{-2}$,
whereas $\hat{S} \simeq (\pi/G)[2/(3H^2)+4/M^2] \approx 4\pi/(GM^2)$ 
during inflation ($H^2 \gg M^2$).
Hence the horizon entropy $S$ in the equilibrium framework 
increases faster than $\hat{S}$ in the non-equilibrium one, 
as in the models with $(1+\sqrt{3})/2<n<2$. 
This property is clearly seen in the numerical simulations 
of Fig.~\ref{inflationfig} ($0<Mt<30$).

As long as $H^2 \gg M^2$ both $\rd S$ and the last term
on the r.h.s. of Eq.~(\ref{usere})
are approximately given by $\pi M^2/(3GH^3) \rd t$. 
Meanwhile we have $(1/F)\rd \hat{S}=
\pi M^4/(18GH^5) \rd t$, which is suppressed by a factor 
of $M^2/(6H^2)$ relative to $\rd S$.
Hence the variation of the horizon entropy $S$ is mainly 
sourced by the entropy production term $\rd_i \hat{S}$ 
during inflation.

The inflationary period is followed by a reheating phase in which 
the Ricci scalar $R$ exhibits a damped oscillation 
with a frequency $M$ [see Eq.~(\ref{steq2})].
The evolution of the Hubble parameter during the reheating period 
can be estimated as \cite{Mijic}
\begin{eqnarray}
H  &\simeq& \biggl[ \frac{3}{M}+\frac34 (t-t_{\rm os})
+\frac{3}{4M} \sin M(t-t_{\rm os}) \biggr]^{-1} \nonumber \\
&& \times \cos^2 \left[ \frac{M}{2}(t-t_{\rm os}) \right]\,,
\end{eqnarray}
where $t_{\rm os}$ is the time at which $H$ starts to oscillate.
Taking the time average of the oscillations in the region 
$M(t-t_{\rm os}) \gg 1$ it follows that 
$\langle H \rangle \simeq (2/3)(t-t_{\rm os})^{-1}$
and hence the Universe evolves as a matter-dominated one 
($a \propto (t-t_{\rm os})^{2/3}$).

The quantity $F=1+R/(3M^2)$ approaches 1 after the end of inflation. 
Equations (\ref{frire1d}) and (\ref{frire2d}) correspond to the 
standard Friedmann equations during the radiation and matter eras. 
$S$ approaches $\hat{S}$ 
after the end of inflation ($F \simeq 1$), which can be  
confirmed in the numerical simulation of Fig.~\ref{inflationfig}. 
In this regime the entropy production term 
$
\left(1/F\right) \left[ \left(2-\beta\right)/\left(4-\beta\right) \right] 
\rd_i \hat{S}
$ 
can be negligible relative to $\rd S$, so that $\rd S \simeq \rd \hat{S}$. 
There are intervals in which the horizon entropies decrease
because of the oscillation of $H$, but the important point 
is that both $S$ and $\hat{S}$ globally increase
in proportion to $\langle H \rangle^{-2} \propto (t-t_{\rm os})^2$.

The entropy production term $\rd_i \hat{S}$ is a dominant 
contribution to $\rd S$ during inflation, but after inflation 
it begins to be suppressed relative to $\rd S$ by a factor
of the order $\langle H \rangle^2/M^2~(\ll 1)$.
This shows that it is more convenient to take the equilibrium 
framework in terms of the single horizon entropy $S$ rather than 
the non-equilibrium framework that separates the horizon 
entropy into two contributions. 

\subsection{Dark energy}

Let us next proceed to $f(R)$ dark energy models 
consistent with both cosmological and local gravity constraints.
We focus on models in which cosmological solutions have a late-time
de Sitter attractor at $R=R_1~(>0)$ satisfying the condition
$Rf_{,R}=2f$. For the stability of the de Sitter point 
we require that \cite{Sch,Faraoni,AGPT} 
\begin{equation}
0<m(R_1)<1\,,\qquad m(R) \equiv \frac{Rf_{,RR}}{f_{,R}}\,,
\label{stacon}
\end{equation}
where $f_{,RR} \equiv \rd^2 f/\rd R^2$. 
Here the quantity $m$ characterizes the deviation from 
the $\Lambda$CDM model ($f(R)=R-2\Lambda$).

There are several other conditions that viable $f(R)$
dark energy models need to satisfy:
(i) $f_{,R}>0$ for $R \ge R_1$ to avoid ghosts, 
(ii) $f_{,RR}>0$ for $R \ge R_1$ to ensure the stability 
of cosmological perturbations \cite{cosmoper}
and to realize a matter-dominated
epoch followed by the late-time cosmic acceleration \cite{APT},
(iii) $m$ rapidly approaches $+0$ for $R \gg R_0$
($R_0$ is the cosmological Ricci scalar today) 
to satisfy local gravity constraints \cite{lgcpapers}. 
In other words the models need to be close to 
the $\Lambda$CDM model
in the region $R \gg R_0$.
More precisely, we require that $m(R) \lesssim 10^{-15}$
for $R \approx 10^{5}R_0$ \cite{Tsuji08,Tsuji08d}.

A number of authors proposed viable models consistent with 
the above requirements \cite{AGPT,Li,AT08,Hu,Star,Appleby,Tsuji,NO07,Linder}.
One representative model is \cite{Star}
\begin{equation}
f(R)=R-\lambda R_c \left[ 
1-\left(1+\frac{R^2}{R_c^2}\right)^{-n} \right]\,,
\label{stamodel}
\end{equation}
where $\lambda$, $R_c$, and $n$ are positive constants.
For $\lambda={\cal O}(1)$, $R_c$ is of the order of $R_0$.
Another similar model is 
$f(R)=R-\lambda R_c (R/R_c)^{2n}/[(R/R_c)^{2n}+1]$ \cite{Hu}, 
which has the same asymptotic form 
$f(R) \simeq R-\lambda R_c [1-(R/R_c)^{-2n}]$
as that in the model (\ref{stamodel}).
When $n>0.9$ these models are consistent with local 
gravity constraints due to the rapid decrease of $m$ 
for increasing $R$ \cite{CT}.

A simpler $f(R)$ model that has only two free parameters
$\lambda$ and $R_c$ is \cite{Tsuji}
\begin{equation}
f(R)=R-\lambda R_c \tanh \left(\frac{R}{R_c}\right)\,, 
\label{fRtanh}
\end{equation}
in which $m \simeq 8 \lambda (R/R_c)e^{-2R/R_c}$ in the region
$R \gg R_c$. For increasing $R$ the quantity $m$ approaches $+0$ 
even faster than in the model (\ref{stamodel}).
Another similar model is 
$f(R)=R-\lambda R_c (1-e^{-R/R_c})$ \cite{Linder}, in which case
$m \simeq \lambda (R/R_c)e^{-R/R_c}$ for $R \gg R_c$.

%%%%%%%%%%%%%%
\begin{figure}
\includegraphics[height=3.4in,width=3.4in]{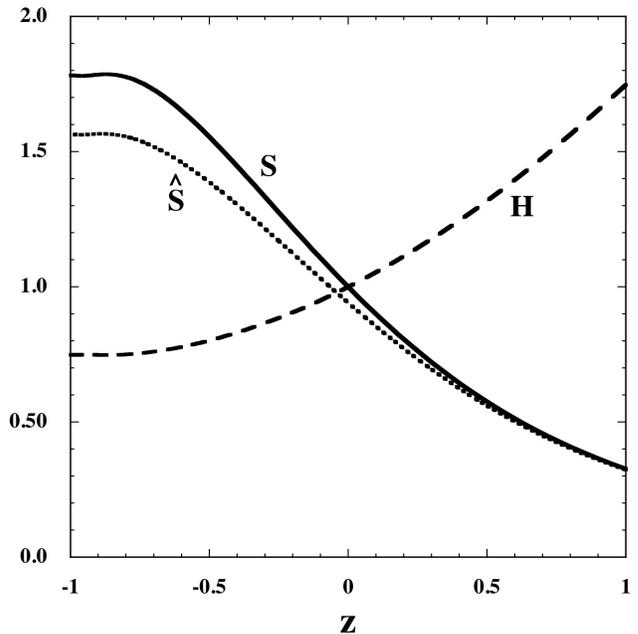}
\caption{\label{defig} 
The evolution of the horizon entropies $S=A/(4G)$ and 
$\hat{S}=FA/(4G)$ versus the redshift $z$
for the dark energy model (\ref{stamodel}) 
with model parameters $n=1$ and $\lambda=1.8$.
The entropy $S$ is normalized to be 1 today ($z=0$).
The initial evolution of $S$ is similar to that of $\hat{S}$, 
but the difference appears as the quantity $F$ 
deviates from 1.}
\end{figure}
%%%%%%%%%%%%%%

For the viable $f(R)$ models mentioned above the quantity 
$F$ is close to 1 in the region $R \gg R_c$, 
so that the evolution of the horizon entropy $S$ is 
similar to that of $\hat{S}$ for the redshift 
$z \equiv a_0/a-1 \gg 1$ ($a_0$ is the scale factor today).
The deviation from the $\Lambda$CDM model appears
for low redshifts ($z \lesssim 1$), which leads to the 
difference between $S$ and $\hat{S}$.
Since $f_{,RR}>0$ for $R \ge R_1$, we have $\dot{F}<0$
and hence $F<1$ provided that $R$ decreases 
with time ($\dot{R}<0$).
This means that $\hat{S}$ should be smaller than $S$
for low redshifts, which can be confirmed in the 
numerical simulation of Fig.~\ref{defig} for the 
model (\ref{stamodel}). 

In the following we study the evolution of $S$ and $\hat{S}$
in more details. The horizon entropy $S \propto H^{-2}$ 
in the equilibrium picture increases as long as $H$ continues 
to decrease toward the de Sitter attractor.
The stability of the de Sitter point given in Eq.~(\ref{stacon})
can be divided into two cases: 
(a) stable spiral for $0<m(R_1)<16/25$ and 
(b) stable node for $16/25<m(R_1)<1$ \cite{AGPT}.\footnote{
This comes from the fact that the eigenvalues for the $3 \times 3$
matrix of perturbations about the de Sitter point are given by $-3$, 
$-3/2 \pm \sqrt{25-16/m(R_1)}/2$.}

In the case (a) the solutions approach the attractor
with the oscillation of $R$, whereas 
in the case (b) the oscillation of the Ricci scalar
does not occur around $R=R_1$.
In the former case the horizon entropy $S$
finally approaches a constant value at the de Sitter point 
with small oscillations. The numerical simulation in 
Fig.~\ref{defig} corresponds to this situation with 
$m(R_1)=0.358$, which shows that 
the oscillation of $S$ around the redshift $-1<z<-0.8$ 
is really tiny.
In the case (b) we have numerically checked that
such oscillations of $S$ disappear, as expected analytically.
We have carried out numerical simulations for other 
viable $f(R)$ models such as (\ref{fRtanh}) and found 
that the above properties also persist in those models. 
Thus the horizon entropy $S$ globally increases with time apart from 
small oscillations that appear for the case   
$0<m(R_1)<16/25$.

Let us estimate the contribution of $\rd \hat{S}$ and 
$\rd_i \hat{S}$ to $\rd S$ in Eq.~(\ref{usere}).
We shall consider the cosmological epoch in which 
the quantity $\dot{H}/H^2$ is approximately 
constant, that is, $\dot{H}/H^2 \simeq -(3/2)(1+w_{\rm eff})$, 
where $w_{\rm eff}$ is an effective equation of the system
($w_{\rm eff}=1/3, 0, -1$ during radiation, matter, and de Sitter
eras, respectively).
Then the last term in Eq.~(\ref{usere}) is approximately 
given by 
\begin{equation}
\overline{\rd_i \hat{S}} \equiv \frac{1}{F}
\frac{2H^2+\dot{H}}{4H^2+\dot{H}}\rd_i \hat{S}
\simeq -\frac{2 \pi}{G} \frac{\dot{H}}{H^3}
\frac{Rf_{,RR}}{f_{,R}} \rd t\,.
\end{equation}
Using $\rd S=-(2\pi/G)(\dot{H}/H^3)\rd t$, the entropy 
production term can be simply expressed as
\begin{equation}
\overline{\rd_i \hat{S}} \simeq m\,\rd S\,.
\label{entfR}
\end{equation}
It then follows from Eq.~(\ref{usere}) that 
\begin{equation}
\rd \hat{S}/F \simeq (1-m)\,\rd S\,. 
\label{entfR2}
\end{equation}
As long as $m \ll 1$ the entropy production 
term in Eq.~(\ref{entfR}) is negligible relative to $\rd S$, 
so that $\rd S \simeq (1/F)\rd \hat{S}$.
As the deviation from the $\Lambda$CDM model 
appears. i.e. $m \gtrsim {\cal O}(0.1)$, 
the entropy production provides an important 
contribution to $\rd S$.

As long as the stability condition (\ref{stacon}) 
is satisfied, the deviation parameter has been 
in the range $0<m<1$ until the solutions reach the de Sitter attractor.
Provided that $\rd S>0$, we then have $\rd_i \hat{S}>0$
and $\rd \hat{S}>0$ from Eqs.~(\ref{entfR}) and (\ref{entfR2}).
In fact the growth of $\hat{S}$ can be confirmed in Fig.~\ref{defig}, 
apart from the tiny oscillations around the de Sitter attractor. 
%%%%%%%%%%%%%%%%%%%%

%%%%%%%%%%%%%%%%%%%
\section{Conclusions}
%%%%%%%%%%%%%%%%%%%

In the present paper, we have studied thermodynamics on the apparent 
horizon with area $A$ in the FLRW space-time for modified gravity theories 
with the Lagrangian density $f(R, \phi, X)$. 
If we define the energy momentum tensor of ``dark'' components
other than perfect fluids as Eq.~(\ref{Tmunud}) with the 
Einstein equation (\ref{Ein1}), the corresponding energy density 
(\ref{rodef1}) and the pressure (\ref{pdef1}) do not satisfy the standard
continuity equation for the theories in which the quantity 
$F=\partial f/\partial R$ is not constant.
Introducing the Wald's horizon entropy in the form $\hat{S}=AF/(4G)$
associated with a Noether charge,  
we have derived the first-law of thermodynamics 
given by Eq.~(\ref{noneqfirst}) in the presence of a non-equilibrium entropy 
production term $\rd_i \hat{S}$.
This non-equilibrium picture of thermodynamics arises for the 
theories with $\rd F \neq 0$, which include $f(R)$ gravity and 
scalar-tensor theories.

If we define the energy density $\rho_d$ and the pressure $P_d$
of dark components as in Eqs.~(\ref{rodef1d}) and (\ref{pdef1d})
respectively, we obtain the standard continuity equation (\ref{conser}).
This corresponds to the introduction of the energy momentum tensor 
$T_{\mu \nu}^{(d)}$ as Eq.~(\ref{eneequ})
with the Einstein equation (\ref{redeein}).
Introducing the Bekenstein-Hawking entropy in the form $S=A/(4G)$,
we have found that the first law of equilibrium thermodynamics
follows from Einstein equations.
Note that this is different from the approach taken in Ref.~\cite{Gong}
for the realization of equilibrium thermodynamics
in which a modified Misner-Sharp mass ${\cal M}$ was introduced
with the definition of the horizon entropy $\hat{S}=AF/(4G)$.

The horizon entropy $S$ in our equilibrium framework is analogous to 
that in Einstein gravity.
We note, however, that this equilibrium thermodynamics in the Jordan 
frame is not in general identical to that in the Einstein frame.
For scalar-tensor theories with the action (\ref{stensor})
we have derived the explicit relations between thermodynamical
quantities in two frames. 
Our equilibrium description of thermodynamics in the Jordan frame 
is valid even for modified gravity theories in which the
Einstein frame action does not exist.
In addition the Jordan frame should be regarded as 
a physical one because of the conservation law of baryons.
Hence the direct construction of equilibrium thermodynamics 
in the Jordan frame makes much more sense 
relative to that in the Einstein frame.

In the flat FLRW background the horizon entropy $S$ 
is proportional to $H^{-2}$, 
which grows for decreasing $H$.
In other words the violation of the null energy condition ($\rho_T+P_T \ge 0$, 
where $\rho_T$ and $P_T$ are the total energy density and the pressure 
respectively) can lead to the decrease of $S$.
We have applied our formalism to $f(R)$ inflation models 
with the Lagrangian density $f(R)=R+\alpha R^n$ with $(\sqrt{3}+1)/2<n \le 2$
and showed that $S$ globally increases apart from the oscillation in the 
reheating phase. The global increase of $S$ also persists in $f(R)$ dark energy 
models that satisfy cosmological and local gravity constraints in which 
the solutions approach a de Sitter attractor.

As we have derived in Eq.~(\ref{usere}), the variation of $S$ can be 
expressed in terms of $\rd \hat{S}$ in the non-equilibrium framework 
together with the entropy production term $\rd_i \hat{S}$.
The last term in Eq.~(\ref{usere}), which we denote $\overline{\rd_i \hat{S}}$, 
can be important for the models in which the quantity $F$ departs from 1.
In $f(R)$ dark energy models, for example, it follows that 
$\overline{\rd_i \hat{S}} \simeq m\,\rd S$ and 
$\rd \hat{S}/F \simeq (1-m)\rd S$, 
where $m=Rf_{,RR}/f_{,R}$
corresponds to the deviation parameter from the $\Lambda$CDM model.
As long as $m \ll 1$ the entropy production term can be negligible 
such that $\rd S \simeq \rd \hat{S}/F$, but its contribution
to $\rd S$ becomes important for $m \gtrsim {\cal O}(0.1)$.
The transition from the former to the latter regime indeed 
occurs at low redshifts for viable $f(R)$ dark energy models.

We have thus shown that the equilibrium description of thermodynamics
in the Jordan frame is present for general modified gravity theories.
The equilibrium description of thermodynamics
is useful not only to provide the General Relativistic analogue
of the horizon entropy irrespective of gravitational theories 
but also to understand the nonequilibrium thermodynamics
deeper in connection with the standard equilibrium framework.
It will be of interest to apply our formalism to dark energy dominated 
universe by taking into account the entropies of dark energy as well as
matter inside the horizon along the lines of Ref.~\cite{Wu2}.

%%%%%%%%%%%%%%%%%%%%%%%%%%%%%%%%
\section*{ACKNOWLEDGEMENTS}
%%%%%%%%%%%%%%%%%%%%%%%%%%%%%%%%

K.B. acknowledges the KEK theory exchange program 
for physicists in Taiwan and the very kind hospitality at 
KEK and  Tokyo University of Science. 
S.T. thanks for the warm hospitality at National Tsing Hua
University where the present work was initiated.
The work by K.B. and C.Q.G. is supported in part by 
the National Science Council of R.O.C. under: 
Grant \#s: NSC-95-2112-M-007-059-MY3 and
NSC-98-2112-M-007-008-MY3
and National Tsing Hua University under the Boost Program and Grant \#: 
97N2309F1.  
S.T. thanks financial support for the Grant-in-Aid 
for Scientific Research Fund of the JSPS (No.~30318802)
and the Grant-in-Aid for Scientific Research on Innovative 
Areas (No.~21111006).

%%%%%%%%%%%%%%%%%%%%%%%%%%%%%%%%%
%% thebibliography environment
%%%%%%%%%%%%%%%%%%%%%%%%%%%%%%%%%


\begin{thebibliography}{99}

\bibitem{Beken}
J.~D.~Bekenstein,
%``Black holes and entropy,''
Phys.\ Rev.\  D {\bf 7}, 2333 (1973).

\bibitem{Bardeen}
J.~M.~Bardeen, B.~Carter and S.~W.~Hawking,
%``The Four laws of black hole mechanics,''
Commun.\ Math.\ Phys.\  {\bf 31}, 161 (1973).

\bibitem{Hawking}
S.~W.~Hawking,
%``Particle Creation By Black Holes,''
Commun.\ Math.\ Phys.\  {\bf 43}, 199 (1975).

\bibitem{Jacobson}
T.~Jacobson,
%``Thermodynamics of space-time: The Einstein equation of state,''
Phys.\ Rev.\ Lett.\  {\bf 75}, 1260  (1995).

\bibitem{Frolov}
A.~V.~Frolov and L.~Kofman,
%``Inflation and de Sitter thermodynamics,''
JCAP {\bf 0305}, 009 (2003).

\bibitem{Danielsson}
U.~H.~Danielsson,
%``Transplanckian energy production and slow roll inflation,''
Phys.\ Rev.\  D {\bf 71}, 023516 (2005). 

\bibitem{Bousso}
R.~Bousso,
%``Cosmology and the S-matrix,''
Phys.\ Rev.\  D {\bf 71}, 064024 (2005).

\bibitem{Hayward}
S.~A.~Hayward, 
%``General laws of black hole dynamics,''
Phys.\ Rev.\  D {\bf 49}, 6467 (1994);
%``Unified first law of black-hole dynamics and relativistic thermodynamics,''
Class.\ Quant.\ Grav.\  {\bf 15}, 3147 (1998);
S.~A.~Hayward, S.~Mukohyama and M.~C.~Ashworth,
%``Dynamic black-hole entropy,''
Phys.\ Lett.\  A {\bf 256}, 347 (1999).

\bibitem{Paddy}
T.~Padmanabhan,
%``Classical and quantum thermodynamics of horizons 
%in spherically  symmetric spacetimes,''
Class.\ Quant.\ Grav.\  {\bf 19}, 5387 (2002).

\bibitem{Akbar}
M.~Akbar and R.~G.~Cai,
%``Thermodynamic Behavior of Friedmann Equation at Apparent Horizon of FRW
%Universe,''
Phys.\ Rev.\  D {\bf 75}, 084003 (2007).

\bibitem{Eling}
C.~Eling, R.~Guedens and T.~Jacobson,
%``Non-equilibrium Thermodynamics of Space-time,''
Phys.\ Rev.\ Lett.\  {\bf 96}, 121301  (2006). 

\bibitem{Akbar2}
M.~Akbar and R.~G.~Cai,
%``Friedmann equations of FRW universe in scalar-tensor gravity, f(R)  gravity
%and first law of thermodynamics,''
Phys.\ Lett.\  B {\bf 635}, 7 (2006);
%``Thermodynamic Behavior of Field Equations for f(R) Gravity,''
Phys.\ Lett.\  B {\bf 648}, 243 (2007).

\bibitem{Gong}
Y.~Gong and A.~Wang,
%``The Friedmann equations and thermodynamics of apparent horizons,''
Phys.\ Rev.\ Lett.\  {\bf 99}, 211301 (2007). 

\bibitem{Wu1}
S.~F.~Wu, B.~Wang and G.~H.~Yang, 
%``Thermodynamics on the apparent horizon in generalized gravity theories,''
Nucl.\ Phys.\  B {\bf 799}, 330  (2008). 

\bibitem{Wu2}
S.~F.~Wu, B.~Wang, G.~H.~Yang and P.~M.~Zhang, 
%``The generalized second law of thermodynamics in generalized gravity
%theories,''
Class.\ Quant.\ Grav.\  {\bf 25}, 235018  (2008). 

\bibitem{Bamba09}
K.~Bamba and C.~Q.~Geng,
%``Thermodynamics in $F(R)$ gravity with phantom crossing,''
Phys.\ Lett.\  B{\bf 679}, 282 (2009). 

\bibitem{Cai07}
R.~G.~Cai and L.~M.~Cao,
%``Unified first law and thermodynamics of apparent horizon in FRW
%universe,''
Phys.\ Rev.\  D{\bf 75}, 064008 (2007).

\bibitem{Paddy2}
A.~Paranjape, S.~Sarkar and T.~Padmanabhan,
%``Thermodynamic route to field equations in Lancos-Lovelock gravity,''
Phys.\ Rev.\  D {\bf 74}, 104015 (2006).

\bibitem{CaiCao08}
R.~G.~Cai, L.~M.~Cao, Y.~P.~Hu and S.~P.~Kim, 
%``Generalized Vaidya Spacetime in Lovelock Gravity and Thermodynamics on
%Apparent Horizon,''
Phys.\ Rev.\  D {\bf 78}, 124012 (2008).

\bibitem{braneth}
A.~Sheykhi, B.~Wang and R.~G.~Cai,
%``Thermodynamical Properties of Apparent Horizon 
%in Warped DGP Braneworld,''
Nucl.\ Phys.\  B {\bf 779}, 1 (2007);
Phys.\ Rev.\  D {\bf 76}, 023515 (2007);
X.~H.~Ge,
%``First law of thermodynamics and Friedmann-like equations in braneworld
%cosmology,''
Phys.\ Lett.\  B {\bf 651}, 49 (2007);
S.~F.~Wu, G.~H.~Yang and P.~M.~Zhang,
%``Cosmological equations and Thermodynamics on 
%Apparent Horizon in Thick Braneworld,''
arXiv:0710.5394 [hep-th].

\bibitem{Cai:2005ra}
R.~G.~Cai and S.~P.~Kim,
%``First law of thermodynamics and Friedmann equations of
%Friedmann-Robertson-Walker universe,''
JHEP {\bf 0502}, 050 (2005). 

\bibitem{Wald1}
R.~M.~Wald,
%``Black hole entropy is the Noether charge,''
Phys.\ Rev.\  D {\bf 48}, 3427 (1993). 

\bibitem{Wald2}
V.~Iyer and R.~M.~Wald,
%``Some properties of Noether charge and a proposal for dynamical black hole
%entropy,''
Phys.\ Rev.\  D {\bf 50}, 846 (1994). 

\bibitem{Hwang}
J.~c.~Hwang and H.~Noh,
%``Classical evolution and quantum generation in generalized gravity  theories
%including string corrections and tachyon: Unified analyses,''
Phys.\ Rev.\  D {\bf 71}, 063536 (2005);
S.~Tsujikawa,
%``Matter density perturbations and effective gravitational constant in
%modified gravity models of dark energy,''
Phys.\ Rev.\  D {\bf 76}, 023514 (2007).

\bibitem{Misner}
C.~W.~Misner and D.~H.~Sharp,
%``Relativistic equations for adiabatic, spherically symmetric gravitational
%collapse,''
Phys.\ Rev.\  {\bf 136}, B571 (1964).

\bibitem{Brustein}
R.~Brustein, D.~Gorbonos and M.~Hadad,
%``Wald's entropy is equal to a quarter of the horizon area in units of the
%effective gravitational coupling,''
Phys.\ Rev.\  D {\bf 79}, 044025 (2009). 

\bibitem{Maeda}
K.~i.~Maeda,
%``Towards the Einstein-Hilbert Action via Conformal Transformation,''
Phys.\ Rev.\  D {\bf 39}, 3159 (1989).

\bibitem{Tsuji08}
S.~Tsujikawa, K.~Uddin, S.~Mizuno, R.~Tavakol and J.~Yokoyama,
%``Constraints on scalar-tensor models of dark energy from 
%observational and local gravity tests,''
Phys.\ Rev.\  D {\bf 77}, 103009 (2008).

\bibitem{Amencoupled}
L.~Amendola,
%``Coupled quintessence,''
Phys.\ Rev.\  D {\bf 62}, 043511 (2000).

\bibitem{Star80}
A.~A.~Starobinsky,
%``A new type of isotropic cosmological models without singularity,''
Phys.\ Lett.\  B {\bf 91}, 99 (1980).

\bibitem{Mijic}
M.~B.~Mijic, M.~S.~Morris and W.~M.~Suen,
%``The R**2 Cosmology: Inflation Without A Phase Transition,''
Phys.\ Rev.\  D {\bf 34}, 2934 (1986).

\bibitem{Sch}
V.~Muller, H.~J.~Schmidt and A.~A.~Starobinsky,
%``The Stability Of The De Sitter Space-Time In Fourth Order Gravity,''
Phys.\ Lett.\  B {\bf 202}, 198 (1988).

\bibitem{Faraoni}
V.~Faraoni,
%``The stability of modified gravity models,''
Phys.\ Rev.\  D {\bf 72}, 124005 (2005).

\bibitem{AGPT}
L.~Amendola, R.~Gannouji, D.~Polarski and S.~Tsujikawa,
%``Conditions for the cosmological viability of f(R) dark energy models,''
Phys.\ Rev.\  D {\bf 75}, 083504 (2007).

\bibitem{cosmoper}
S.~M.~Carroll, I.~Sawicki, A.~Silvestri and M.~Trodden,
%``Modified-Source Gravity and Cosmological Structure Formation,''
New J.\ Phys.\  \textbf{8}, 323 (2006);
T.~Faulkner, M.~Tegmark, E.~F.~Bunn and Y.~Mao,
%``Constraining f(R) gravity as a scalar tensor theory,''
Phys.\ Rev.\  D {\bf 76}, 063505 (2007);
Y.~S.~Song, W.~Hu and I.~Sawicki,
%``The large scale structure of f(R) gravity,''
Phys.\ Rev.\  D {\bf 75}, 044004 (2007);
R.~Bean, D.~Bernat, L.~Pogosian, A.~Silvestri and M.~Trodden,
%``Dynamics of Linear Perturbations in f(R) Gravity,''
Phys.\ Rev.\  D \textbf{75}, 064020 (2007);
Y.~S.~Song, H.~Peiris and W.~Hu,
%``Cosmological Constraints on f(R) Acceleration Models,''
Phys.\ Rev.\  D {\bf 76}, 063517 (2007);
L.~Pogosian and A.~Silvestri,
%``The pattern of growth in viable f(R) cosmologies,''
Phys.\ Rev.\  D {\bf 77}, 023503 (2008).

\bibitem{APT}
L.~Amendola, D.~Polarski and S.~Tsujikawa,
%``Are f(R) dark energy models cosmologically viable ?,''
Phys.\ Rev.\ Lett.\  {\bf 98}, 131302 (2007);
Int.\ J.\ Mod.\ Phys.\  D {\bf 16}, 1555 (2007).

\bibitem{lgcpapers}
G.~J.~Olmo,
Phys.\ Rev.\  D {\bf 72}, 083505 (2005);
A.~L.~Erickcek, T.~L.~Smith and M.~Kamionkowski,
Phys.\ Rev.\  D {\bf 74}, 121501 (2006);
V.~Faraoni,
Phys.\ Rev.\  D {\bf 74}, 023529 (2006);
I.~Navarro and K.~Van Acoleyen,
%``f(R) actions, cosmic acceleration and local tests of gravity,''
JCAP {\bf 0702}, 022 (2007);
T.~Chiba, T.~L.~Smith and A.~L.~Erickcek,
%``Solar System constraints to general f(R) gravity,''
Phys.\ Rev.\  D {\bf 75}, 124014 (2007);
P.~Brax, C.~van de Bruck, A.~C.~Davis and D.~J.~Shaw,
%``f(R) Gravity and Chameleon Theories,''
Phys.\ Rev.\  D {\bf 78}, 104021 (2008);
I.~Thongkool, M.~Sami, R.~Gannouji and S.~Jhingan,
%``The generosity of $f(R)$ gravity models 
%with disappearing cosmological constant,''
Phys.\ Rev.\  D {\bf 80}, 043523 (2009).

\bibitem{Tsuji08d}
S.~Tsujikawa, R.~Gannouji, B.~Moraes and D.~Polarski,
%``The dispersion of growth of matter perturbations in f(R) gravity,'' 
Phys.\ Rev.\  D {\bf 80}, 084044 (2009). 

\bibitem{Li}
B.~Li and J.~D.~Barrow,
%``The Cosmology of f(R) Gravity in the Metric Variational Approach,''
Phys.\ Rev.\  D {\bf 75}, 084010 (2007).

\bibitem{AT08}
L.~Amendola and S.~Tsujikawa,
%``Phantom crossing, equation-of-state singularities, and local gravity
%constraints in $f(R)$ models,''
Phys.\ Lett.\  B {\bf 660}, 125 (2008).

\bibitem{Hu}
W.~Hu and I.~Sawicki,
%``Models of f(R) Cosmic Acceleration that Evade Solar-System Tests,''
Phys.\ Rev.\  D {\bf 76}, 064004 (2007).

\bibitem{Star}
A.~A.~Starobinsky,
%``Disappearing cosmological constant in f(R) gravity,''
JETP Lett.\  {\bf 86}, 157 (2007).

\bibitem{Appleby}
S.~A.~Appleby and R.~A.~Battye,
%``Do consistent $F(R)$ models mimic General Relativity plus $\Lambda$?,''
Phys.\ Lett.\  B {\bf 654}, 7 (2007).

\bibitem{Tsuji}
S.~Tsujikawa,
%``Observational signatures of f(R) dark energy models 
%that satisfy cosmological and local gravity constraints,''
Phys.\ Rev.\  D {\bf 77}, 023507 (2008).

\bibitem{NO07}
S.~Nojiri and S.~D.~Odintsov,
Phys.\ Lett.\  B {\bf 657}, 238 (2007).

\bibitem{Linder}
E.~V.~Linder,
%``Exponential Gravity,''
Phys.\ Rev.\  D {\bf 80}, 123528 (2009). 

\bibitem{CT}
S.~Capozziello and S.~Tsujikawa,
%``Solar system and equivalence principle constraints 
%on $f(R)$ gravity by chameleon approach,''
Phys.\ Rev.\  D {\bf 77}, 107501 (2008). 

\end{thebibliography}
\end{document}